\documentclass[aps,prd,tighteness,preprint,superscriptaddress,showpacs]{revtex4}
\usepackage{epsf,epsfig,graphics,graphicx}
\usepackage{verbatim,color,ulem}
\newcommand{\be}{\begin{equation}}
\newcommand{\ee}{\end{equation}}
\newcommand{\bea}{\begin{eqnarray}}
\newcommand{\eea}{\end{eqnarray}}
\newcommand{\ba}{\begin{array}}
\newcommand{\ea}{\end{array}}
\usepackage{amssymb,amsmath,amsthm}
\usepackage[mathscr]{eucal}

\begin{document}

\title{Time evolution of entanglement entropy of moving mirrors influenced by strongly coupled quantum critical fields}
\author{Da-Shin Lee}
\email{dslee@mail.ndhu.edu.tw} \affiliation{Department of Physics,
National Dong-Hwa University, Hualien, Taiwan, R.O.C.}
\author{Chen-Pin Yeh}
\email{chenpinyeh@mail.ndhu.edu.tw} \affiliation{Department of
Physics, National Dong-Hwa University, Hualien, Taiwan, R.O.C.}

\begin{abstract}
The evolution of the Von Neumann entanglement entropy of a
$n$-dimensional mirror influenced by the strongly coupled
$d$-dimensional quantum critical fields with a dynamic exponent
$z$ is studied by the holographic approach. The dual description
is a $n+1$-dimensional probe brane moving in the $d+1$-dimensional
asymptotic Lifshitz geometry ended at $r=r_b$, which plays a role
as the UV energy cutoff. Using the holographic influence
functional method, we find that in the linear response region, by
introducing a harmonic trap for the mirror, which serves as a IR
energy cutoff, the Von Neumann entropy at late times will saturate
by a power-law in time for generic values of $z$ and $n$. The
saturated value and the relaxation rate depend on the parameter
$\alpha\equiv 1+(n+2)/z$, which is restricted to $1<\alpha <3$ but
$\alpha \ne 2$. We find that the saturated values of the entropy
are qualitatively different for the theories with $1<\alpha<2$ and
$2<\alpha<3$. Additionally, the power law relaxation follows the
rate $\propto t^{-2\alpha-1}$. This probe brane approach provides
an alternative way to study the time evolution of the entanglement
entropy in the linear response region that shows the similar
power-law relaxation behavior as in the studies of entanglement
entropies  based on Ryu-Takayanagi conjecture. We also compare our
results with quantum Brownian motion in a bath of relativistic
free fields.
\end{abstract}


\maketitle
\section{Introduction}
Entanglement entropies provide useful probes to non-local
properties of quantum systems, which are important for
understanding quantum phase transitions\cite{q_pha}, and moreover
are the key quantities in quantum information
processing\cite{q_inf}. The idea of entanglement entropies has
also received much attention in connection to the information
paradox in black hole physics \cite{bh_inf}. It is generally
impossible to isolate a particular quantum system in which we are
interested from its surrounding. Considering a full theory that
describes the interaction between system and environment, from
which all information like correlation functions can in principle
be obtained. Tracing or integrating out the degrees of freedom of
the environment to obtain an effective field theory for the system
leads to a loss of information. If the quantum state in the full
theory is a pure state, namely a zero entropy state, tracing out
the environmental degrees of freedom yields a reduced density
matrix for the degrees of freedom of the system, which typically
becomes a mixed state with non-vanishing entropy. The Von-Neumann
entropy is a measure of the loss of information in the process of
integrating out some degrees of freedom in the full unitary system
\cite{hubook,brbook}. The effective theory can be described by the
reduced density matrix $\rho_r$, which is obtained by tracing out
the environmental variables in the full density matrix using the
method of Feyman-Vernon influence functional\cite{Fv}. The Von
Neumann entropy is then defined by $-Tr\rho_r\ln\rho_r$ with the
trace over system's variables. In general, the effective theory
for the system obtained in this way is not unitary, and it can
lead to the dissipative and stochastic behavior of the system.
This concept has been used in pioneering works on quantum Brownian
motion\cite{Leggett} and general open quantum
systems\cite{hubook}.  In particular, the relaxation of the system
into equilibrium with the environment can be characterized by the
time evolution of the entanglement entropy. The method of
Feynman-Vernon influence functional has also been extended to the
quantum field theory \cite{Leggett,SK}. However, the influence
functional can only be exactly derived if the environmental
theories are Gaussian and their coupling with the system is
linear~\cite{GSI,hubook}. In this paper, we would like to study
the entanglement between a particle or a mirror and some strongly
coupled quantum fields by a holographic construction of the
influence functional that has been proposed in~\cite{Son:2009vu}
and \cite{Yeh_14}.

The idea of holographic duality is originally proposed as the
correspondence between $4$-dimensional conformal field theory and
gravity theory in $5$-dimensional anti-de Sitter
space~\cite{AdSCFT}, and soon is generalized to other backgrounds
and field theories. One aim is to provide a framework to study the
strong coupling problems in the condensed matter systems (see
\cite{Hartnoll_09} for a review). The holographic approach has
also be adapted to tackle the problems of the Brownian motion of a
particle moving in a strongly coupled
environment~\cite{Herzog:2006gh,Gubser_06,Teaney_06,Son:2009vu,Giecold:2009cg,CasalderreySolana:2009rm,Huot_2011,Holographic
QBM,Tong_12,Hartnoll_10,mirror,Yeh_14,Yeh_16_1,Yeh_16_2,Yeh_18_1,Yeh_18_2,GIA}.
The idea is that a particle immersed in a environment given by the
quantum field corresponds to a bulk fundamental string ended at
the boundary of the dual gravity theory. In the black hole
background, the string undergoes random motion due to the Hawking
radiation of the transverse fluctuation modes. This is the bulk
dual of the thermal Brownian motion. A review on the holographic
Brownian motion can be found in \cite{Holographic QBM}. Another
aim of the holographic duality is to understand the quantum
behavior of black hole or quantum gravity in general. A recent
proposal by Ryu and Takayanagi~\cite{Ryu}, that the entanglement
entropy in the boundary theory is related to the minimal area in
the gravity theory, has arisen the hope that quantum gravity may
be formulated in the language of quantum information. Time
evolution of the entanglement entropy has also been studied in
this framework~\cite{Hubeny_07,Hartman_13,Liu_13}, as a measure of
the relaxation rate for non-equilibrium systems and also as a
probe to the black hole interior.

In this work, we apply a bottom-up holographic method, proposed in
our earlier works~\cite{mirror} and \cite{Yeh_14}, to study the
evolution of the entanglement entropy for the system of a
$n$-dimensional mirror in the environment of $d$-dimensional
quantum critical theories with dynamical exponent $z$ at zero
temperature. The holographic dual for such quantum critical
theories has been proposed in~\cite{Kachru_08} where the gravity
theory is in the Lifshitz background
(See~\cite{Tong_12,Hartnoll_10} for details). Several physical
phenomena have been studied in this theory, including linear DC
conductivity, power-law AC conductivity, and strange fermion
behaviors~\cite{Hartnoll_10,Gursoy_12,Alisha,Gursoy,dimer model}.
In our set-up, the bulk counterpart of the mirror is a
$(n+1)$-brane in the Lifshitz geometry in $d+1$ dimensions. The
dynamics of the mirror can be realized from the motion of the
brane ended at the boundary of the bulk at the radius distance
$r_b$. As explained in \cite{mirror} and will also be reviewed in
the appendix, this holographic identification is based upon the
fact that the coupling of the brane to the boundary field shares
similar feature as the coupling between the mirror and the
environment quantum field, where the mirror of perfect reflection
effectively sets the vanishing boundary condition for the
field~\cite{wu,mirror}. In the case of $n=0$, it becomes a
fundamental string in the bulk and its end point describes the
position of a particle in the boundary. Using the method of
holographic influence functional, developed in~\cite{Yeh_14}, we
are able to derive the reduced density matrix of the mirror for a
given initial state. The trick to adopt is that in the linear
response approximation, with an initial Gaussian state of the
system, the reduced density matrix remains Gaussian when turning
on the bilinear couplings between the system and environment, and
is completely determined by the expectation values of the position
operator $\hat{q}$, momentum operator $\hat{p}$ and their product
$\hat{q}\hat{p}+\hat{p}\hat{q}$. Thus the time dependent reduced
density matrix can be constructed from the expectation values with
the influence functional, rather than solving the master equation
for the reduced density matrix.  The evolution of the entanglement
entropy shows how the system equilibrate with the environment and
lose the information about its initial state, as seen from the
increase of the entropy comparing to the entropy in the initial
state of the system.

To summarize, in this paper we consider quantum critical fields
with $1<\alpha<3$ and $\alpha\neq2$  where $\alpha\equiv
1+\frac{n+2}{z}$ that couple to the system of the mirror.  Our
results show that the Von Neumann entanglement entropy of the
system relaxes by the power law in time as $t^{-2\alpha-1}$ toward
saturation. Since the boundary field has the dispersion relation
$E\propto k^z$, in the $d-1$ spatial dimension the density of
state $\rho(E)\propto E^{-1+\frac{d-1}{z}}$. For a $n$-dimensional
mirror under consideration, the effective spatial dimension probed
by the mirror is $n$. Thus, the environmental density of states
available to the mirror would be $\rho (E)\propto
E^{-1+\frac{n}{z}}$. Accordingly, the density of modes decreases
as increasing $z$ or decreasing $n$. If the relaxation of the
system is attributed to dissipation of energy of its degrees of
freedom to the modes of the environment, the relaxation rate
decreases as $\alpha$ is decreased. Moreover, we find that the
saturate value of the entanglement entropy,
$S(t\rightarrow\infty)\simeq
(1-\frac{1}{\alpha})\ln\left(\frac{r_b^z}{\Omega}\right)$ for
$1<\alpha<2$ and $S(t\rightarrow\infty)\simeq
(\frac32-\frac{\alpha}{2})\ln\left(\frac{r_b^z}{\Omega}\right)$
for $2<\alpha<3$, where $\Omega$ is the oscillation frequency of
the harmonic trap for the mirror. The value of $\Omega$, which
serves as the IR cutoff in the energy scale in this system, should
be small enough comparing to $r_b^z$, which is the UV energy
cutoff. This shows the qualitative difference for the theories in
two regions, $1<\alpha<2$ and $2<\alpha<3$. It can be seen that
when other parameters  are fixed, the saturated Von-Newmann
entropy reaches maximal when $\alpha$ approaches $2$. This marked
the transition relates to the fact that in the IR limit the mass
of the mirror is an irrelevant operator as $\alpha < 2$ and a
relevant operator as $\alpha> 2$. However, for $\alpha >3$, it is
also found that to have the Von-Newmann entropy consistent with
the minimum uncertainty relation of quantum mechanical systems
requires the curvature radius $L$ in the Lifshitz metric to be of
the order of Planck length $l_p$. This implies the breakdown of
the assumption in the holographic approach by treating the
background geometry as the classical configuration given by the
solution of Einstein equations. The quantum gravity effects need
to be taken into account in this case and deserves further study.

Our presentation is organized as follows. In next section, we
review the idea of open quantum systems and introduce the method
of the closed-time-path formalism. The environmental degrees of
freedom in the full density matrix of the system-plus-environment
are traced over to obtain the reduced density matrix of the
system. Environmental effects are then all encoded in the
influence functional, which can be completely determined by the
nonequilibrium two-point correlators in the linear response
approximation. In Sec.~\ref{sec2}, we briefly review the method of
the holographic influence functional, and present the analytical
form of the nonequilibrium two-point correlators to be used in the
later calculation. The more detailed derivation of these
correlators is in Appendix. In Sec.~\ref{sec3}, we study the
dynamics of the mirror with a harmonic trap potential, immersed in
the environment of strongly coupled quantum critical fields.
Before turning on the interaction of mirror with the environment,
the initial state is assumed to be the direct product of the
mirror's ground state and the vacuum state of the quantum critical
field. When turning on the interaction, we obtain the effective
nonequilibrium action by tracing out the environment fields and
introducing the noise's degree of freedom. The Heisenberg
equations of motion for the mirror's degree of freedom can be
derived from the effective action. Their real-time solutions can
be studied using the Laplace transform technique. In
Sec.~\ref{sec4}, we obtain the time evolution of the entanglement
entropy from the expectation value of $\hat q^2$, $\hat p^2$, and
$\hat p \hat q+ \hat q \hat p$, and show our main results. In
Sec.~\ref{sec5}, the comparison to the case with the environment
fields given by the relativistic free fields is made in the
paradigm of quantum Brownian motion. Concluding remarks are in
Sec.~\ref{sec6}.

\section{Nonequilibrium effective action and Influence functional} \label{sec1}
In quantum systems, the complete information of the expectation
values and correlation functions can be determined by the time
dependent density matrix $\rho (t)$. The closed-time-path
formalism enables us to calculate the evolution of the density
matrix that has been prepared at some particular initial time
$t_i$. In this work, we consider the system linearly coupled to an
environment field. The full Lagrangian consisting of the
system-plus-environment, takes the form
  \be
   L(q,F)=L_q[q]+L_F[F]+qF\, ,
  \ee
where $q$ and $F$ generically represent the system and the
environment variables respectively. We assume that the initial
density matrix at time $t_i$ can be factorized as
\begin{equation}\label{initialcond}
    \rho(t_i)=\rho_{q}(t_i)\otimes\rho_{{F}}(t_i)\, ,
\end{equation}
where $\rho_{{F}}(t_i)$ is the initial density matrix of the
environment. The full density matrix $\rho(t)$ evolves unitarily
according to
\begin{equation}
 \rho (t_f) = U(t_f, t_i) \, { \rho} (t_i) \, U^{-1} (t_f,
t_i )
\end{equation}
with $ U(t_f,t_i) $ the time evolution operator, involving the
degrees of freedom of the system and environment. The reduced
density matrix $\rho_r$ of the system of interest can be obtained
by tracing over the environmental degrees of freedom, $F$ in the
full density matrix and can be written as~\cite{Leggett,SK,GSI}
\begin{equation}
\rho_r({q}_f,\tilde{{q}}_f,t_f)=\int\!d{q}_1\,d{q}_2\;\mathcal{J}({q}_f,\tilde{{q}}_f,t_f;{q}_1,{q}_2,t_i)\,\rho_{q}({q}_1,{q}_2,t_i)\,,\label{evolveelectron}
\end{equation}
where the propagating function
$\mathcal{J}({q}_f,\tilde{{q}}_f,t_f;{q}_1,{q}_2,t_i)$ is
\begin{equation}\label{propagator}
    \mathcal{J}({q}_f,\tilde{{q}}_f,t_f;{q}_1,{q}_2,t_i)=\int^{{q}_f}_{{q}_1}\!\!\mathcal{D}{q}^+\!\!\int^{\tilde{{q}}_f}_{{q}_2}\!\!\mathcal{D}{q}^-\;\exp\left[i\int_{t_i}^{t_f}dt\left(L_{q}[{q}^+]-L_{q}[{q}^-]\right)\right]\mathcal{F}[{q}^+,{q}^-]\, ,\end{equation}
with a path integral that propagates forward and backward
in time. For the environment of a free field theory or in the
linear response approximation, the influence functional
$\mathcal{F}[{q}^+,{q}^-]$ can be written in terms of real-time
Green's functions of the environment fields $F$~\cite{Fv},
\bea\label{influencefun2}
{\mathcal{F}}\left[{q}^{+},{q}^{-}\right]&& = \exp\bigg\{
-\frac{i}{2}\int_{t_i}^{t_f} dt\!\!\int_{t_i}^{t_f} \!dt' \Big[
{q}^+(t)\,G^{++}(t,t') \,{q}^+(t')\Bigr.\bigr.
-{q}^+(t)\,G^{+-}(t,t')\,{q}^-(t') \nonumber\\
&&- {q}^-(t)\,G^{-+}(t,t')\,{q}^+(t')
\big.\big.+{q}^-(t)\,G^{--}(t,t')\,{q}^-(t')\big]\bigg\}\,.
 \eea
The Green's functions involved are time-ordered, anti-time-ordered
and Wightman functions, defined as \bea \label{correlator}
    && i\,G^{+-}(t,t')=\langle F(t')F(t)\rangle \, , \nonumber\\
    && i\,G^{-+}(t,t')=\langle F(t)F(t')\rangle \, ,\nonumber\\
    &&i\,G^{++}(t,t')=\langle F(t)F(t')\rangle\theta(t-t')+\langle
    F(t')F(t)\rangle\theta(t'-t)\, , \nonumber\\
    && i\,G^{--}(t,t')=\langle F(t')F(t)\rangle\theta(t-t')+\langle
    F(t)F(t')\rangle\theta(t'-t) \, ,
    \eea
where the expectation values are calculated in the state
described by $\rho_F(t_i)$. The retarded Green's function and
Hadamard function can be constructed from them according to \bea
\label{G_HR}
    G_{R} (t-t')&\equiv & -i \theta (t-t')\langle [F(t), F(t')] \rangle = \bigg\{ G^{++} ( t,t') -G^{+-} (t,t') \bigg\}\,,  \\
    G_{H} (t-t')&\equiv & \frac{1}{2} \langle \{ F(t), F(t') \} \rangle=\frac{i}{4} \bigg\{ G^{++} ( t,t') +G^{+-} (t,t') + G^{--} ( t,t') +G^{-+} (t,t')
    \bigg\}\,. \nonumber
    \eea
In a time-translation invariant environment, the Fourier transform
of various Green's functions is defined by
\begin{equation} G (t-t')=\int \, \frac{ d \omega}{2\pi} \, G
(\omega) \, e^{-i \omega (t-t')} \, . \end{equation}

Notice that the above Green's functions are not totally independent as a result of  the unitarity property of the system-plus-environment. In particular, as the environment is in thermal
equilibrium at the temperature $T$ initially with
$\rho_{F}(t_i)=e^{-\frac1{T}H_F}$, the fluctuation-dissipation
relation gives \be \label{FD} G_H(\omega)=- (1+2n_{\omega}) \,
{\rm Im} G_R (\omega) \, \ee with
$n_{\omega}=(e^{\frac{\omega}{T}}-1)^{-1}$. In the next section,
we briefly review the construction of Lifshitz geometry and the method of
holographic influence functional, with which we explore the
effects of the strongly coupled quantum critical fields on the
system.

\section{The holography model and Holographic influence functional} \label{sec2}
The environment that we will consider is described by the theory
of $d$-dimensional quantum critical points with the following
scaling symmetry:
  \be\label{E:ernfd}
  t\rightarrow\mu^zt \, ,\qquad\qquad x\rightarrow\mu x \,
  \ee
where $z$ is called the dynamical exponent. It's gravity dual is
described by the d+1-dimensional Lifshitz geometry with the
metric~\cite{Kachru_08},
   \be
   \label{bmetric}
   ds^2=-\frac{r^{2z}}{L^{2z}}dt^2+\frac{1}{r^2}dr^2+\frac{r^2}{L^2}dx_idx_{i}
   \, ,
   \ee
where the scaling symmetry {\eqref{E:ernfd}} is realized as an
isometry of this metric. In the following calculations we will set
the curvature radius $L=1$, and restore $L$ when giving the order
of magnitude estimate on the quantities of the system. This
$d$+1-dimensional Lifshitz metric can be constructed by coupling
gravity with negative cosmological constant to massive Abelian
vector fields \cite{Marika_08}. The influence functional for the
mirror ($n\ge 1$) can be obtained in the gravity theory by
computing the on-shell DBI action of the brane. The details can be
found in~\cite{Yeh_14,Son_09} and also in the Appendix. There we
first introduce $Q^+(t,r_1)$ and $Q^-(t,r_2)$, which describe the
brane's positions around the stationary configuration in two
regions with different asymptotic boundaries in the maximally
extended Lifshitz black hole geometry~\cite{Son_09, Son_02}. We
then impose the boundary conditions at $r_b$
 \be
 \label{bc} {q^{\pm}(t)=Q^{\pm}(t,r_{b})}\,,
 \ee
and the analyticity conditions in the black hole horizons. By
identifying the variables $q^{\pm} (t)$ as the displacement of the
moving mirror in the close-time-path formalism, the classical
on-shell action of the brane can then be identified as the
influence functional for the mirror~\cite{Son_09}:
   \be
   \label{gravity action}
   \mathcal{F} [q^+,
   q^-]= S_{gravity}\left(Q^+(t,r),Q^-(t,r)\right) =S^{\rm on-shell}_{DBI}(Q^+)-S^{\rm on-shell}_{DBI}(Q^-)\,
   \ee
where $S^{\rm on-shell}_{DBI}$ is the on-shell DBI action for the
probe brane. We call this construction of $\mathcal{F} [q^+,q^-]$
the holographic influence functional.  Various
Green's functions of the field can be read off (See Appendix).

The DBI action in the case of the $n+1$-dimensional probe brane in the
Lifshitz black hole, up to the quadratic order in perturbations, is written in Appendix.
Here we just present it in the zero temperature limit ($r_h \rightarrow 0 $)
for the sake of introducing the notations and the assumptions,
 given by \be
  S_{DBI}
  \simeq {\rm constant}- \frac{T_{n+1}}{2}\int dr \, dt \, dx_1 \, dx_2 \, ... \, dx_n \,
\bigg( r^{z+n+3}  X'^{I} X'^{I}-
\frac{\dot{X}^{I}\dot{X}^{I}}{ {  r^{z-n-1}}}\bigg) \, ,
  \label{s_T_dbi_1}
  \ee
where $T_{n+1}$ is the tension of the brane and $X^I(t,r)$
parameterizes the brane's position around the stationary
configuration, where $I=n+1,...,d$ denotes the transverse
directions to the brane. Also, $X'^{I} =\partial_{r} X^{I}$,
$\dot{X}^{I}=\partial_{t}X^{I}$. We assume that the mirror does
not deform when moving in its transverse directions so that
all $X^{I}$ depends only on $t$ and $r$.  In the quadratic order, the
perturbations in different directions decouple, and a particular direction $X^{I}$ is considered.
Again, we start from the DBI action for the probe brane in the
Lifshitz black hole, from which to be able to determine $Q^{\pm}$ in two outside regions of the black hole separately. In the zero temperature limit by taking $r_h\rightarrow 0$, the zero-temperature retarded Green's
function  with $\alpha=1+(n+2)/z$, for $\omega>0$ can be found to be,
  \be
\label{G_R} G_R(\omega)=- T_{n+1} S_n \,  {\omega \,
r_b^{n+2}}\frac{H^{(1)}_{\frac{\alpha}{2}-1}(\frac{\omega}{zr_b^z})}{H^{(1)}_{\frac{\alpha}{2}}(\frac{\omega}{zr_b^z})}
\,,
  \ee
where $H^{(1)}_{\nu}(x)$ is the Hankel
function of the first kind and $S_n$ is the volume of the mirror.
As we will particularly focus on the late time dynamics, the small $\omega$
expansion of $G_R(\omega)$ is considered,
 \be \label{G_R_approx}
  G_R (\omega)={ m (i\omega)^2+\gamma (-i\omega)^{\alpha}}+ {\cal{O}}( \omega^2/r_b^{2z}) \,
  ,
  \ee
 where
  \be \label{m_mu_n}
   m =\frac{T_{n+1} S_n}{z(\alpha-2)r_b^{z(2-\alpha)}},
  \ee
and
 \be \label{gamman} \gamma =\frac{T_{n+1}
S_n}{(2z)^{\alpha-1}}\frac{\Gamma(1-\frac{\alpha}{2})}{\Gamma(\frac{\alpha}{2})}
\, .
 \ee
In this expansion there are the terms of even powers in $\omega$
to be treated as the correction to the dispersion relation. We
regard the $\omega^{\alpha}$ term as the leading contribution to
the damping effect. Apparently, the interaction between the mirror
and quantum critical fields not only gives the mirror's mass  $ m$
with the $r_b$ dependence but also induce the damping effect on
the mirror with the coefficient $\gamma $.  The friction
coefficient $\gamma$ is independent of $r_b$, which becomes
important in stabilizing the dynamics of the mirror in a
fluctuating environment. The zero-temperature Hadamard function
for $\omega
>0$ is found to be,
 \be \label{G_0_H}
 G_H(\omega)=\frac{2 z}{\pi} r_b^{n+2+z} \frac{T_{n+1} S_n} {J^2_{\frac{\alpha}{2}}(\frac{\omega}{zr_b^z})+Y^2_{\frac{\alpha}{2}}(\frac{\omega}{zr_b^z})}
  \, , \ee
where $J_{\nu}(x)$ and $Y_{\nu}(x)$ are Bessel functions. Also, in
the small $\omega$ limit, the Hadamard function can be
approximated by \be \label{G_0_H_approx}
 G_H(\omega)=\frac{\pi \, T_{n+1} S_n }{ (2 z)^{\alpha-1} \, \Gamma^2 (\frac{\alpha}{2})} \, \omega^{\alpha} + {\cal{O}}( \omega^{\alpha+2}/r_b^{z(\alpha+2)}) \, .
\ee Notice that the obtained retarded Green's function and the Hadamard
function obey the zero-temperature fluctuation-dissipation
theorem.

\section{Dynamics of a moving mirror} \label{sec3}
Let us now specify the Lagrangian of the system.  We see from the
previous section that the interaction with the environment field
gives  the mass correction $m$ (\ref{m_mu_n}) to the mirror in the small $\omega$ limit.
We here introduce  the bare mass $m_0$ where the renormalized
mass becomes $M=m+m_0$. To give the system
an infrared energy scale, we also add a quadratic trap potential with the oscillation frequency $\Omega_0$  to the
mirror.
Thus, the Lagrangian for the mirror with the displacement
$q$ can be written as
\be \label{mirror_L}
 L_q[q]=\frac{1}{2} m_0 \dot{q}^2 -\frac{1}{2} m_0
\Omega_0^2 q^2 \, . \ee
The frequency $\Omega_0$ is set to be small comparing to the UV
cutoff $r_b^z$, and plays a role as an IR cutoff. The retarded
Green's function in~(\ref{G_R}) after subtracting the mass
correction term $m$ is defined to be the self-energy
$\Sigma(\omega)$
 \be \label{G_R_m} \Sigma (\omega)=
G_R (\omega) -m (i\omega)^2 \, .
 \ee

To study the stochastic dynamics of the system, it is more
convenient to change the $q^{+}$, $q^{-}$ coordinates to the
average and relative coordinates:
\begin{equation}
 q= (  q^+ + q^- )/2 \, , \,\,\,\,\,
q_{\Delta}= q^+ - q^- \, .
\end{equation}
Thus, the coarse-grained effective action can be defined
from~(\ref{propagator}) by using the Lagrangian of the mirror in
(\ref{mirror_L}) and the holographic influence functional
in~(\ref{influencefun2}), as
 \bea
 & & S_{\rm CG} \big[q^{\pm} = q \pm \frac{q_{\Delta}}2 \big]  = \int_{t_i}^{t_f} dt ~( L_q[q^+]-L_q[q^-] ) -i   \,  \ln{\cal
F} \left[  q^+,  q^- \right] \nonumber\\
&& \quad =  \int_{t_i}^{t_f}  dt  \int_{t_i}^{t_f}  dt'   (M
\dot{q}_{\Delta}(t) \dot{q}(t') - M \Omega^2   {q}_{\Delta}(t)
{q}(t') )  -   \int_{t_i}^{t_f} dt   \int_{t_i}^{t} dt'
q_{\Delta}(t) \Sigma (t-t')
q(t') \nonumber \\
&& \quad\quad\quad + \frac{i}{2}  \int_{t_i}^{t_f}  dt
\int_{t_i}^{t_f} dt' q_{\Delta}(t) \, G_H (t-t')  q_{\Delta} (t')
\, .
\eea
where $G_H(t)$ and $\Sigma(t)$ are the inverse Fourier
transform of $G_H(\omega)$ and $\Sigma(\omega)$, and
$\Omega=(1-\frac{m}{M})\Omega_0$. Apparently, the relevant Green's
functions that contribute to the mirror's dynamics at quadratic
order are the retarded Green's function and Hadamard function,
which we have obtained above in the holographic setup. After
tracing over the degrees of freedom of the environment, the system
of the interest becomes non-unitary, and the resulting
coarse-grained effective action becomes complex-valued function.
Before turning on the interaction with the environment, the
initial density matrix of the mirror is assumed to be in its
ground state of a simple harmonic oscillator with frequency
$\Omega$,
\be \label{initial_d} \rho_q (q, q'; t_i)=  \bigg( \frac{ M
\Omega}{\pi} \bigg)^{1/2} e^{- [ M \Omega^2 ( q^2+q'^2)]} \, , \ee
where the initial position and momentum uncertainties are given
respectively by $ (\Delta q)^2 (t_i)=(2 M \Omega)^{-1}$ and
$(\Delta p )^2 (t_i) =\frac12 M \Omega$. It is possible to carry
out the path integral over the mirror's degree of freedom
in~(\ref{propagator}) and derive the reduced density of the matrix
at time $t$. Then, the entanglement entropy of the system can be
computed by taking the trace of the reduced density matrix with
respect to the degrees of freedom of the mirror via $S=-Tr_q
[\rho_r \ln \rho_r]$. Instead of resorting to this straightforward
but cumbersome method, we use the fact that, in the linear
response approximation, the resulting reduced density matrix, with
the initial condition in (\ref{initial_d}), remains Gaussian when
the system is linearly coupled to the environment \footnote{We
thank Jen-Tsung Hsiang for pointing out this trick to us}. Then
the entanglement entropy can also be computed from the expectation
values of $ \hat q^2 , \hat p^2 $, and $ \hat q \hat p + \hat p
\hat q $, where $\hat q$ and $\hat p$ are respectively the
position and the momentum operators of the
mirror~\cite{Joos_85}. To find the time dependent expectation
values, we need to derive the equation of motion for the position
operator, incorporating the effects from the environment. This can
be done by introducing an auxiliary variable $ \eta (t)$, the
noise force, with a Gaussian distribution function:
\begin{equation}
P[\eta (t)] = \exp \left\{ - \frac{1}{2}  \, \int_{t_i}^{t_f} dt
\, \int_{t_i}^{t_f} dt' \, \eta (t) \, G^{ -1}_{H} (t-t') \, \eta
(t') \right\} \, . \label{noisedistri}
\end{equation}
In terms of the noise force $\eta (t) $,  $ S_{CG} $ can be
rewritten as an ensemble average over $\eta (t)$,
\begin{equation}
\exp i S_{CG}  =  \int {\cal D} \eta \, P [\eta (t)] \, \exp i
S_{\eta} \left[ q, q_{\Delta} ; \eta \right] \, ,
\end{equation}
where the stochastic coarse-grained effective action   $ S_{\eta
}$
is given by
\bea
S_{\eta} [q, q_{\Delta}  ; \eta ] && = \int_{t_i}^{t_f}  dt  \int_{t_i}^{t_f}  dt'   (M \dot{q}_{\Delta}(t) \dot{q}(t') - M \Omega^2   {q}_{\Delta}(t) {q}(t') ) \nonumber \\
&&   -   \int_{t_i}^{t_f} dt   \int_{t_i}^{t} dt'   q_{\Delta}(t)
\Sigma (t-t') q(t') +  \int_{t_i}^{t_f} \eta(t)  q_{\Delta}(t) \,
.
\eea
Then, in the canonical formalism, the equation of motion of the
position operator is found to be,
\be M \, \ddot{\hat q}(t) + M \, \Omega^2 \hat q(t) +\int_{0}^t
\Sigma (t-t') \, \hat{q}(t') dt' = \hat \eta (t) \,
\label{langevin} \ee
with the initial time set to be $t_i=0$. There are two different
sources of quantum fluctuations, over which the averages are taken
on  mirror's position. One is the average over intrinsic quantum
fluctuations of the mirror, and the other is the average over the
noise manifested from quantum fluctuations of the environment.
The noise distribution function $P[\eta(t)]$ in (\ref{noisedistri}) leads to the
correlation function of the noise as follows
\be \langle \hat \eta (t) \, \hat \eta(t') \rangle= G_H (t-t') \,.
\ee
For the environment consisting of a collection of harmonic
oscillators, the Heisenberg equations of motion of the system and
the environment can be solved exactly, giving the same form of the above equation with the associated
 Green's functions constructed out of the harmonic
oscillators~\cite{hubook,brbook,boyan}.  In this work, the
influence functional is derived holographically, and in this way
the retarded Green's function and Hadamard function contains the
information of the strongly coupled quantum critical fields.

The solutions of (\ref{langevin}) can be expressed in term of the
fundamental solutions $G_1 (t)$ and $G_2 (t)$, defined
to be the solutions of the homogeneous equations without the noise term, obeying the following boundary
conditions~\cite{GSI,boyan}:
\bea
&& G_1 (0)=1 \, \, , \dot{G}_1 (0)=0 \, ; \label{G1ini} \\
&& G_2 (0)=0 \, \, , \dot{G}_2 (0)=1 \, . \label{G2ini}
\eea
The Laplace transform of them is respectively defined as
\be
G_{1,2} (s)
=\int_0^{\infty} G_{1,2} (t) \,\, e^{-s t} dt\, .
\ee
Then $G_{1,2} (s)$ can be found to be
\bea
G_1 (s) &&= \frac{s}{ s^2 +\Omega^2 + \tilde \Sigma (s)} \, ; \nonumber \\
G_2 (s) &&= \frac{1}{ s^2 +\Omega^2 + \tilde  \Sigma (s)} \,
\label{Gs}
\eea
with the function $\tilde\Sigma(s)\equiv \Sigma( -i s)/M$, where
$\Sigma(\omega)$ is the self-energy of the mirror~(\ref{G_R_m}).
The solution of $\hat{q}(t)$ can be expressed in terms of
$G_1(t)$ and $G_2 (t)$ as
\be \label{hat q} \hat{q}(t) =G_1 (t) \, \hat{q}(0) + G_2 (t) \,
\frac{\hat{p}(0)}{M} +\frac{1}{M} \int_0^t dt' \, G_2 (t-t') \,
\hat \eta(t') \, . \ee
The real-time function of $G_{1,2}$ is
given by the inverse Laplace transform
\be G_{1,2} (t) = \frac{1}{2 \pi i} \int_{C} \, e^{ s t}\, G_{1,2}
(s) ds \, , \label{invLaplace} \ee where $C$ refers to the
Browmwich contour running along the imaginary axis to the right of
all the poles and cuts of $G_{1,2} (s)$ in the complex s plane.
Therefore we need to understand the analytical structure of the
self-energy $\Sigma$ to obtain the real-time dynamics of all
expectation value of the mirror's variables. Here we will mainly
focus on the late time dynamics of the mirror to find its
relaxation dynamics as well as saturation behavior on the lose of
information about the initial conditions of the system influenced
by the environment. Accordingly, the retarded Green's function and
Hadamard function in their small-$\omega$ approximation
in~(\ref{G_R_approx}) and (\ref{G_0_H_approx}) will be applied. In
principle, we can study the early time behavior since the exact
form of the Green's functions is known. However, due to the
complicated analytical structure of the self energy, we will leave
it for the future study.

\begin{figure}
\centering \scalebox{0.6}{\includegraphics[trim=2cm 4cm 2cm
2cm,clip]{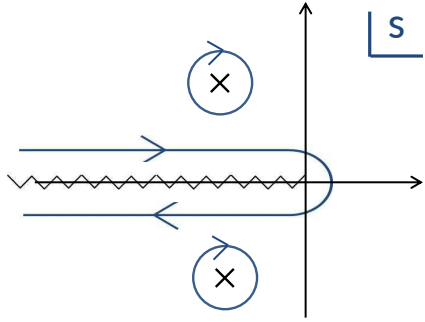}} \caption{The contour for the inverse Laplace
transform used to compute the fundamental solutions $G_1(t)$ and
$G_2(t)$ with the existence of the cut and the poles.}
\label{contour}
\end{figure}

From the expression~(\ref{G_R_approx}) for the small-$\omega$
approximation of the retarded Green's function, the function
$\tilde \Sigma(s)$ is approximated by

\be \tilde \Sigma (s) \simeq \frac{\gamma}{M} s^{\alpha} \, .
\label{selfenergy_app} \ee
For non-integer values of $\alpha$, we choose a branch cut on the
negative real $s$ axis. This amounts to choosing that
\be
\tilde \Sigma (s=-s_0 \pm i
\epsilon)= { {\rm Re}} \tilde \Sigma (-s_0) \pm i { {\rm Im}}
\tilde \Sigma (-s_0) \, ,
\ee
for $s_0 >0$, where
{we define}
\bea
{{\rm  Re}} \tilde \Sigma (-s_0) &&= \frac{\gamma}{M} s_0^{\alpha} \cos(\alpha \pi )\, , \label{SigmaRe}\\
{ {\rm Im}} \tilde \Sigma (-s_0)   &&= \frac{\gamma}{M}
s_0^{\alpha} \sin (\alpha \pi) \, . \label{SigmaIm} \eea
Notice that the same form of the self-energy has also been studied
in the works~\cite{nagy,Ali} where the system is coupled to the
bath of harmonic oscillators. The properties of the bath is
completely characterized by the spectral density that has the
form, for example, $J(\omega) \propto \omega^{\beta}$ for $
\omega>0$ with non-integer $\beta$, where $\omega$ is the frequency of oscillators. Also
notice that the branch cuts arise due to the fact that the
parameter $\alpha=1+(n+2)/z$ is a non-integer. Apart from the
cuts, there exist the poles determined by
\be \label{poleeq} s_p^2+\Omega^2+{\tilde \Sigma(s_p)}=0 \, , \ee
where the real and imaginary parts of $s_p$ correspond to the
oscillatory frequency and the decay rate, and we denote them by
 \be
s_p=\pm i \omega_p -\Gamma \, .
 \ee
Thus these states of the mirror are not exactly the eigenstates of
the interacting Hamiltonian. To have the analytical expressions,
we consider two separate limiting situations with regard to the
relative importance of the kinetic term $s^2$ to the damping term
{$\Sigma(s)\simeq \frac{\gamma}{M} s^{\alpha}$} at the positions
of the poles. These two limiting cases are characterized by the
parameter, $\Delta \equiv \frac{\gamma}{M} \Omega^{\alpha-2} $,
where $|\Delta| <<1 $ is for the kinetic term dominance and
$|\Delta|>>1 $ is for the damping term dominance. Using
(\ref{gamman}), the parameter $\Delta$ can also be written as,
 \be
 \label{delta}
 \Delta=\frac{T_{n+1}\pi \Omega^{\alpha-2}}{(2z)^{\alpha-1}\rho_n \Gamma^2(\frac{\alpha}{2})\sin\frac{\pi\alpha}{2}}
 \ee
with the mass density of the mirror, $\rho_n=M/S_n$. As we will
see later, the runaway solutions with negative $\Gamma$ exist for
$\alpha>2$. From the field theory perspective, the problem of the
existence of the runaway solutions is due to the fact that the
resulting equation involves higher than 2nd-order time derivative
terms, giving the unpleasant feature of instability. For example,
the Abraham-Lorentz equation, which includes the radiation
reaction force for a nonrelativistic charged particle, is
third-order in time and possesses runaway solutions. In
electrodynamics, the problem can be solved \cite{Jackson} by
reformulating the theory to eliminate unstable solutions with the
method of reduction of order that reduces the equation to the
2nd-order. The similar procedure also works in the Einstein
gravity \cite{Ward}. Here we assume all the runway solutions can
be removed by the above mentioned method and discuss stable
solutions only. In case that there are more than one  stable mode,
we concentrate on the most long-lived mode to be discussed in the
following section. Furthermore, as $\alpha>3$, we will see the
breakdown of our assumption of the Gaussian reduced density matrix
from the obtained saturated entanglement entropy, which shows the
mass density of the mirror needs to be of the Planck scale order,
being consistent with the minimum uncertainty of quantum systems.
Thus in the following discussion, we will focus on the damping
term dominated region, $1<\alpha<2$ and the kinetic term dominated
region, $2<\alpha<3$ respectively. In these cases, the Browmwich
contour for the inverse Laplace transform in (\ref{invLaplace})
can be deformed as in Fig. (\ref{contour}). We will discuss the
contributions from the poles and the branch cut separately.

\subsection{the pole contributions}
We now restore the curvature radius $L$ for the order-of-magnitude
estimation of relevant quantities. Then the parameter $\Delta
\simeq (T_{n+1}L^{n+2})(L^{n+1}\rho_n)^{-1}(L\Omega)^{\alpha-2}$.
In the top-down constructions, the tension of the branes, which is
related to the string coupling, determines the Planck length scale
$l_p$ in the gravity theory by $(T_{n+1}L^{n+2})\simeq
\big(\frac{L}{l_p}\big)^{n+2}$. In the classical gravity limit
(corresponding to the large-$N$ limit in the boundary theory), we
should have $(T_{n+1}L^{n+2})\gg 1$. Also, in the boundary theory, we
may choose the unit such that $\rho_n\simeq L^{-n-1}$. If so, the
condition $\Delta\gg 1$ for the damping term dominated can be
achieved for all values of $\alpha$ by choosing $\Omega\simeq
L^{-1}$. As for  $\Delta\ll 1$ in the case of the kinetic term dominated, the choice of $\Omega$ has to satisfy
$(L\Omega)^{\alpha-2}\ll \big(\frac{l_p}{L}\big)^{n+2}$.  In the
following, we will discuss the late time behavior of the mirror in
these two limits separately.

\subsubsection{the damping term dominated region ($\Delta \gg 1$)} \label{damp_dominated}
As $\Delta\gg 1$, the resonance frequency $\omega_p$ and the width
$\Gamma$ can be found  for $1<\alpha<2$ to be
   \bea
\label{omega_gamma_1}
  \omega_p && \simeq \Omega\Delta^{-\frac1{\alpha}}  \,  \sin\frac{\pi}{\alpha}  \, , \nonumber \\
  \Gamma  && \simeq -\Omega\Delta^{-\frac1{\alpha}} \,   \cos\frac{\pi}{\alpha} \, .
   \eea
The width $\Gamma$ is positive for $1<\alpha<2$, corresponding to
the stable solution. Moreover they are the only poles in the
principal sheet. As for $2<\alpha<3$, there are more than two
poles on the principal sheet and also the runaway solutions with
$\Gamma<1$ may exist.  We will assume that these unstable modes
can be removed by the reduction of order method as described in
the previous section, and concentrate on the stable mode. In this
case, the stable mode is found  to be
   \bea
\label{omega_gamma_1}
  \omega_p && \simeq \Omega(-\Delta)^{-\frac1{\alpha}}  \,  \sin\frac{2\pi}{\alpha}  \, , \nonumber \\
  \Gamma  && \simeq -\Omega(-\Delta)^{-\frac1{\alpha}} \,   \cos\frac{2\pi}{\alpha} \, .
   \eea
These poles contribute to the functions $ G_{1,2} (t)$ by
\bea
G_{2, pole} (t) && \propto  \cos [\omega_p t+\varphi] e^{-\Gamma t} \, , \label{G2pole} \\
G_{1, pole} (t) && =  \frac{d}{d t}G_{2, pole} (t) \, ,
\label{G1pole} \eea
where $\varphi$ is a constant. This solution is valid when $t\gg
r_b^{-z}$. Since the resonance frequency and the width are of the
same order, it leads to the
broad resonance behavior as in contrast with the narrow resonance cases
that will occur in the kinetic term dominated region.

\subsubsection{the kinetic term dominated region ($\Delta\ll 1$) } \label{kin_dominated}
As $\Delta\ll 1$, we find the narrow resonance modes of the long-lived resonances,  which  correspond to an
almost  energy eigenstate of the interactive Hamiltonian with
the width much smaller than the oscillation frequency.
The frequency and the width are found perturbatively to be
\be  \label{omega_0} \omega_p = \Omega +\delta \Omega
\, , \ee where
\be \delta \Omega=\frac12\Omega\Delta \cos\bigg(\frac{\alpha
\pi}{2}\bigg) \, ; \ee
and
\be \label{gamma_0} \Gamma  =\frac12\Omega\Delta \sin
\bigg(\frac{\alpha \pi}{2}\bigg)  \,
 \ee
with $\Gamma >0$ in both the ranges, $1<\alpha<2$ and
$2<\alpha<3$. However as $2<\alpha<3$, there are runaway solutions
in the principal sheet, which are presumably removed. No matter it
is narrow or broad resonance, the pole contributions to the
functions $G_1(t)$ and $G_2(t)$ decay exponentially in general as
in (\ref{G2pole}). In the following, we consider the cut
contributions and discuss the time scales, after which $G_1(t)$
and $G_2(t)$ will relax in power-law behavior in stated.

\subsection{the cut contributions}
The cut contributions  to the $G_{1,2}$ functions  can be expressed
as
\bea
G_{2, cut} (t) && = -\frac1{\pi} \int_0^{\infty} d s \,  \frac{ {\rm Im} \tilde{\Sigma} (-s)}{[ s^2+\Omega^2 + {\rm Re} \tilde{\Sigma} (-s)]^2 + [{\rm Im} \tilde{\Sigma}(-s)]^2}\, e^{-s t} \, ,  \label{G1cut_approx} \\
G_{1, cut} (t) && = \frac{d}{dt} G_{2, cut} (t) \, ,
\label{G2cut_approx}
\eea
where ${\rm Re} \tilde{\Sigma} $, and
${\rm Im} \tilde{\Sigma}$ are defined in~(\ref{SigmaRe}) and
(\ref{SigmaIm}) respectively. Notice that in obtaining this
expression, we have already used the small $\omega$ approximation
($\frac{\omega}{r_b^z}\ll 1$) for $\Sigma(\omega)$, to focus on
the late time behavior, namely $t\gg r_b^{-z}$.

Since the integral for the cut contributions is dominated by the
small $s$ region in the late times, simpler results can be
obtained for  much latter times, $t\gg \Omega^{-1}$, comparing to
the late time region ($t\gg r_b^{-z}$), which are the time scales
for  the pole contributions dominated. In this case, the integrand
behaves like $s^{\alpha}e^{-st}$ and the integral can be evaluated
by the Gamma function to be
\be
\label{G2cut_approx2}
G_{2,cut}(t)\simeq
\frac{\Delta}{\pi \Omega}\Gamma(1+\alpha)\sin(\alpha\pi)(\Omega t)^{-1-\alpha} \,.
\ee
The function $G_{1,cut}(t)$ is just the time derivative of
$G_{2,cut}(t)$. Naively, it may look like that the cut
contributions dominate over the pole contributions as
$t>\Gamma^{-1}$, which is the decay time.
However, after comparing $G_{2, cut}(t)$ with $G_{2, pole}(t)$, we
find that in fact for the power-law
behavior of cut contributions to dominate, the relevant time
scales are $t \gtrsim t_n\equiv
-\Omega^{-1}(\alpha+1)\Delta^{-1}\ln \Delta $ for $\Delta\ll 1$
(narrow resonance) and $t \gtrsim t_b\equiv
\Omega^{-1}(1+1/\alpha)\Delta^{1/\alpha}\ln \Delta $ for
$\Delta\gg 1$ (broad resonance).

\begin{figure}[h]
        \includegraphics[scale=0.4]{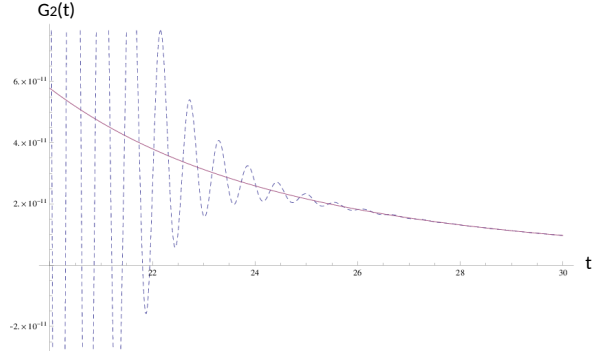}
        \caption{The dash line shows the numerical result
for $G_2(t)$ from the poles and cut contributions with their
expressions (See main text for details) by setting $\alpha=2.4$,
$\Omega=10$, $\frac{\gamma}{M}=-0.01$, and $T_{n+1}S_n=15$. We
also set $\Delta =\frac{\gamma}{M} \Omega^{\alpha-2}=-0.25$, which
corresponds to the case of narrow resonance.  The solid line is
also for the evolution of $G_2(t)$ in the late times ($ t \gg
1/\Omega$) given by (\ref{G2cut_approx2}) with the  same
parameters above.}
        \label{G2_2}
  \end{figure}

\begin{figure}[h]
        \includegraphics[scale=0.4]{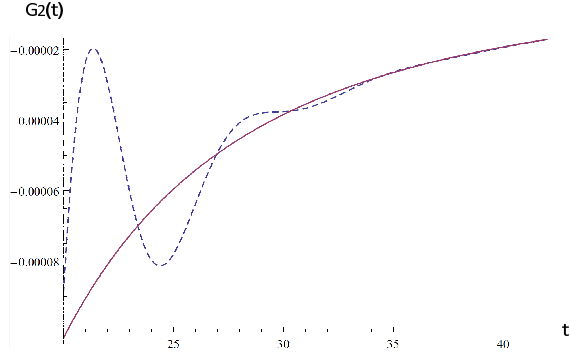}
        \caption{The dash line shows the numerical result
for $G_2(t)$ from the poles and cut contributions with their
expressions (See main text for details) by setting $\alpha=1.4$,
$\Omega=1.5$, $\frac{\gamma}{M}=1.8$, and $T_{n+1}S_n=15$.  We
also set  $\Delta =\frac{\gamma}{M} \Omega^{\alpha-2}=2.2$, which
corresponds to the case of broad resonance.  The solid line is
also for the evolution of $G_2(t)$ in the late times ($ t \gg
1/\Omega$) given by (\ref{G2cut_approx2}) with the  same
parameters above.}
        \label{G2_3}
  \end{figure}

In the following, we numerically check that the function $G_2(t)$
approaches $G_{2,cut}(t)$  found in (\ref{G2cut_approx2}).
The numerics is done by doing $s$-integral in~(\ref{G1cut_approx})
with ${\rm Re} \tilde{\Sigma} (-s)$ and ${\rm Im} \tilde{\Sigma}
(-s)$ as given in~(\ref{SigmaRe}) and (\ref{SigmaIm}), and add the
pole contributions given by (\ref{G2pole}). The results are shown in
Fig.(\ref{G2_2}) with the chosen parameters for narrow resonance.
It can be seen that  $ G_{2}(t)$ is oscillatory in time with the
frequency $\omega_p$ and its amplitude decays exponentially at a
rate determined by $\Gamma$.  After the time  $ t \simeq t_n $,
the $G_2(t)$ function becomes the power-law decay given by
(\ref{G2cut_approx2}). It is interesting to contrast with the
broad resonance as shown in Fig. (\ref{G2_3}) by choosing
appropriate parameters. The $G_2(t)$ function in this case,
is also oscillatory with smaller frequency, and after the time $t\simeq
t_b$,  it exhibits the power-law relaxation as given by the cut
contributions. The $G_1(t)$ function can be similarly checked.

\section{quantum uncertainties on  the mirror and the entanglement entropy}\label{sec4}
Before turning on the interaction with the environment, the
initial state of the system is prepared to be in the ground state of
the harmonic oscillator.
 The environment fields are also in zero-temperature ground state.
The full density matrix is the direct product of the states of the system and the environment field, which is a pure state. The
interaction with the environment turns the reduced density matrix
for the system into a mixed state. However in the linear response
approximation and with the bilinear coupling between the system
and the environment, the reduced density matrix for the system
remains Gaussian at all times. If so, the Von-Neumann entanglement
entropy constructed from the Gaussian reduced density matrix can
be written as~\cite{Joos_85}
\be \label{S_vn} S(t) =\bigg(\sqrt{w (t)}+\frac{1}{2} \bigg) \ln
\bigg( \sqrt{w (t)} +\frac{1}{2} \bigg) - \bigg(\sqrt{w
(t)}-\frac{1}{2} \bigg) \ln \bigg( \sqrt{w (t)} -\frac{1}{2}
\bigg) \, . \ee
where the time dependent function, $w(t)$ is defined by the
position and momentum uncertainties and their cross correlation,
\be  \label{w} w(t)\equiv \langle \hat q^2(t) \rangle  \langle
\hat p^2(t) \rangle - \frac{1}{2} \big( \langle \hat q(t) \hat
p(t) + \hat p(t) \hat q(t)\rangle \big)^2 \, . \ee
Thus in the Gaussian state, the entanglement entropy can be
obtained by calculating these three functions.

Using (\ref{hat q}), the
position uncertainty at time $t$ can be expressed as
\bea
\langle \hat q^2 (t) \rangle  && =\langle \hat q^2 (t) \rangle_I + \langle\hat q^2 (t) \rangle_F \nonumber\\
\langle \hat q^2 (t) \rangle_I && =G_1^2 (t) \langle\hat q^2 (0) \rangle + (G^2_2 (t)/M^2) \langle  \hat p^2 (0) \rangle +(G_1 (t) G_2 (t)/M) \langle \hat q(0)\hat p(0)+\hat p(0) \hat q (0) \rangle  \nonumber\\
\langle \hat q^2 (t) \rangle_F  && = \frac{1}{2 M^2} \int_0^t  d
\tau \int_0^t  d \tau' G_2 ( t-\tau) G_2 (t-\tau') \langle \big\{
\hat \eta (\tau), \hat \eta (\tau') \big\} \rangle \, , \eea
where $\langle \hat q^2 (t) \rangle_I$ is the uncertainty due to
the intrinsic quantum fluctuations of the mirror, and thus depends on
the initial position and momentum uncertainties. However $\langle \hat
q^2 (t) \rangle_F$ is the uncertainty induced by the environment
and involves the two-point function of the environment field.
Using the late time expressions for $G_1(t)$ and $G_2(t)$ ($t\gg
t_n$ for narrow resonance and $t\gg t_b$ for
broad resonance) obtained in last section, the various terms in
$\langle\hat q^2 (t) \rangle_I$  show the power-law relaxation:
\bea \label{q2M_late_time}
G_1^2 (t) \langle \hat q^2 (0) \rangle  && \propto t^{-4-2\alpha} \, , \nonumber\\
(G^2_2 (t)/M^2) \langle \hat p^2 (0) \rangle && \propto t^{-2-2\alpha}  \, , \nonumber\\
(G_1 (t) G_2 (t)/M) \langle \hat q(0) \hat p(0)+ \hat p(0) \hat
q(0) \rangle && \propto t^{-3-2\alpha}  \, . \eea
We will see in the following that $\langle \hat q^2 (t) \rangle_I$ depending on the
initial conditions decay more quickly at late times as comparing
to $\langle \hat q^2 (t) \rangle_F$   due to the environment-induced uncertainty.
The term $\langle \hat q^2 (t)
\rangle_F$ can be further expressed by $\langle \hat q^2 (\infty) \rangle_{F} $, the saturated value, and $\langle \hat
q^2 (t) \rangle_{F,{\rm asy}}$ vanishing at $t\rightarrow \infty$ as
\bea
\langle \hat q^2 (t) \rangle_F && =\frac{1}{2 m^2}  \int_0^t  d \tau \int_0^t  d \tau' G_2 ( t-\tau) G_2 (t-\tau') \langle \big\{ \eta (\tau), \eta (\tau') \big \} \rangle \nonumber\\
&& = \frac{2}{m^2} \int_0^\infty \frac{ d \omega}{ 2\pi} \, G_H (\omega) \bigg\{ \bigg[ \int_0^\infty d\tau -\int_t^\infty d \tau \bigg] G_2 (\tau) e^{-i\omega \tau} \, \nonumber\\
&&\quad\quad\quad \quad\quad\quad \quad\quad\quad\quad\quad \bigg[ \int_0^\infty d\tau' -\int_t^\infty d\tau' \bigg] G_2 (\tau') e^{i\omega \tau'}  \bigg\} \nonumber\\
&& \equiv  \langle \hat q^2 (\infty) \rangle_{F} +  \langle \hat
q^2 (t) \rangle_{F,{\rm asy}} \,,
\eea
The saturated value is determined by $G_H (\omega)$
in~(\ref{G_0_H}) and the Fourier transform of $G_2(t)$, which can
be obtained from $G_2(s)$ in~(\ref{Gs}) by substituting $s=i
\omega$ as
\be
G_2 (\omega)= \frac{1}{-\omega^2+\Omega^2 +{\rm Re} \Sigma
(\omega)/M +i {\rm Im} \Sigma (\omega)/M} \, .
\label{G2omega}
\ee
The self-energy $\Sigma(\omega)$ is defined in (\ref{G_R_m}) and for $\Omega\ll
r_b^z$ we have,
\be
{\rm Re} \tilde{\Sigma} (\omega)\equiv{\rm Re} \Sigma
(\omega)/M\simeq \frac{\gamma}{M}\omega^{\alpha} \cos\frac{\pi
\alpha}{2},~~~ {\rm Im} \tilde{\Sigma} (\omega)\equiv{\rm Im}
\Sigma (\omega)/M\simeq \frac{\gamma}{M}\omega^{\alpha}
\sin\frac{\pi \alpha}{2} \, .
\ee
Then we can write the saturation value as
\bea \langle \hat q^2 (\infty) \rangle_{F}
&& = \frac{2}{M^2}  \int_0^{zr_b^z} \frac{ d \omega}{2 \pi} \, G_H (\omega) \vert G_2(\omega) \vert^2 \nonumber\\
&&  = \frac{2}{M^2} \int_0^{zr_b^z} \frac{ d \omega}{ 2\pi}
\frac{G_H (\omega)}{[-\omega^2+\Omega^2 + {\rm Re} \tilde{\Sigma}
(\omega)]^2+[{\rm Im} \tilde{\Sigma}(\omega)]^2} \, .
\label{q2_F_omega}
 \eea
The integral has a UV-cutoff at $\omega_{UV}=zr_b^z$. With the proper rescaling of
$x\equiv\frac{\omega}{zr_b^z}$ and
$\delta\equiv\frac{\Omega}{zr_b^z}$, we first consider
$\delta=0$ in
the limit of $\Omega\ll r_b^z$. Using the exact expression for $\Sigma(\omega)$ in
(\ref{G_R_m}) and $G_H(\omega)$ in (\ref{G_0_H}), it is found that
   \be
   \label{q2_1}
\langle \hat q^2 (\infty)
\rangle_{F,\delta=0}=\frac{r_b^{z(1-\alpha)}}{\pi
  T_{n+1}S_n}\int_0^1 dx \frac{g(x)}{|Bx^2+x f(x)|^2}
   \ee
where
  \be
  f(x)=\frac{H^{(1)}_{\frac{\alpha}{2}-1}(x)}{H^{(1)}_{\frac{\alpha}{2}}(x)},~~~g(x)=\frac{1}{J^2_{\frac{\alpha}{2}}(x)+Y^2_{\frac{\alpha}{2}}(x)},~~~B=\frac{1}{\alpha-2}-\frac{z\rho_n r_b^{z(2-\alpha)}}{T_{n+1}} \, .
  \ee
The above integrant goes like $x^{\alpha-4}$ for small $x$. The
integral (\ref{q2_1}) converges for $\alpha>3$, and thus $\langle
\hat q^2 (\infty) \rangle_{F,\delta=0}$ gives the leading result
to the saturated value $\langle \hat q^2 (\infty) \rangle_{F}$  as
$\delta\rightarrow 0$. However, the integral (\ref{q2_1}) diverges
for $1<\alpha<3$. By keeping  $\delta$ finite but small, the main
contributions to the integral  (\ref{q2_F_omega}) come from the
small $x$ region, which can be approximated   by
\be \label{q2_s2} \langle \hat q^2 (\infty) \rangle_{F} \simeq
\frac{\pi
r_b^{z(\alpha-3)}T_{n+1}}{z^2S_n2^{\alpha}\rho_n^2\Gamma^2(\frac{\alpha}{2})}
H_{\alpha}(\delta) \,
 \ee
with
\be \label{H_delta} H_{\alpha}(\delta)\equiv
\int_0^1dx\frac{x^{\alpha}}{(-x^2+\delta^2)^2+2A\cos(\frac{\pi\alpha}{2})x^{\alpha}(-x^2+\delta^2)+A^2x^{2\alpha}}\,
 \ee
and $A=\frac{\pi\alpha T_{n+1}r_b^{z(\alpha-2)}}{z\rho_n}$  where
the small $\omega$ expressions for $G_H(\omega)$ and
$\Sigma(\omega)$ have been applied. Because $H_{\alpha}(\delta)$
diverges as $\delta \rightarrow 0$ for $1<\alpha<3$, the small
$\delta$  effectively sets the IR cutoff for the integral in
(\ref{H_delta}).  We see that for $1<\alpha<2$, the IR cutoff is
at $x_{IR}^{2\alpha}\simeq \delta^4$, giving  $x_{IR}\simeq
\delta^{2/\alpha}$, and for $2<\alpha<3$,  $x_{IR}^4\simeq
\delta^4$, thus giving $x_{IR}\simeq \delta$.   In the
$\delta\rightarrow 0$ limit, the different infrared behaviors of
the integrand in (\ref{H_delta}) between two ranges of $\alpha$
are due to the fact that the retarded Green's function in
(\ref{G_R_approx}) in the small $\omega$ expansion is dominated by
the mass term for $1<\alpha <2$ as a relevant operator in the IR
limit, but the damping term for $2<\alpha <3$ where the mass term
becomes an irrelevant operator. So, the integral with the
divergent parts only can be estimated as
\begin{align}
 \label{H}
  H_{\alpha}(\delta\rightarrow 0) & \propto \begin{cases}
                                    \displaystyle \,
  \int_{\delta^{2/\alpha}}^{1} dx \frac{x^{\alpha}}{x^{2\alpha}} \propto {\delta^{\frac2{\alpha}-2}}\, , \quad  1< \alpha <2 \, ;& \vspace{9pt}\\
                                    \displaystyle \,
  \int_{\delta}^1 dx \frac{x^{\alpha}}{x^4} \propto {\delta^{\alpha-3}} \, , \quad  2< \alpha <3
                                      \,&
                                \end{cases}
\end{align}
The different behaviors of the position uncertainty in the
different ranges of $\alpha$ lead to distinct saturated values of
the entanglement entropy as we will see in the following.

\begin{figure}[h]
        \includegraphics[scale=0.4]{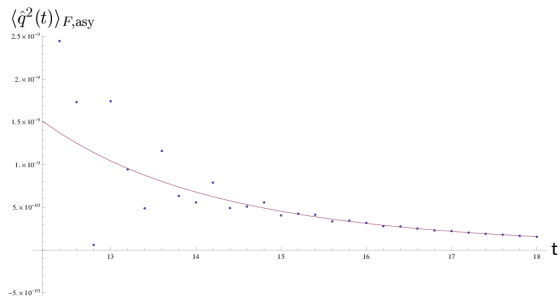}
        \caption{
        The dots show the numerical calculation of $\langle \hat q^2 (t) \rangle_{F,{\rm
        asy}}$,
        using the small $\omega$
        approximate of $G_H$ in~(\ref{G_0_H_approx}) and the $G_2(\omega)$ in (\ref{G2omega}). The solid line
        shows the power law time dependence as we find in
        Eq.(\ref{q2_approx}). In the plots, $\langle \hat q^2 (t) \rangle_{F,{\rm
        asy}}$ is in units of $\frac{2\sin(\alpha\pi)\Gamma(1+\alpha)\sin(\frac{\pi\alpha}{2})\Delta^2}{M\pi\Omega^{2\alpha}}$ with the parameters chosen as in Fig.~\ref{G2_2} for narrow resonance.}
        \label{qq_2}
   \end{figure}

\begin{figure}[h]
        \includegraphics[scale=0.4]{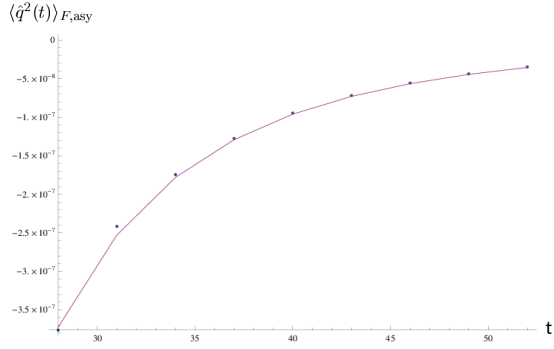}
        \caption{The same numerical calculation as in~Fig.\ref{qq_2} with the parameters
        chosen for the broad resonance as in Fig.\ref{G2_3}.}
        \label{qq_3}
   \end{figure}

At the late-time($t\gg t_n$ ($t\gg t_b$) for narrow (broad)
 resonance), the time dependence of the
position uncertainty is mainly determined by the cut contributions
to $G_2(t)$. Using (\ref{G2cut_approx2}) and the small $\omega$
approximation for $G_2(\omega)$ and $G_H(\omega)$, we obtain the
asymptotic power law behavior as
\bea
 \langle \hat q^2 (t) \rangle_{F,{\rm asy}}  && \simeq-\frac{4}{M^2} \int_0^\infty \frac{ d \omega}{2 \pi} \, G_H (\omega) \, {\rm Re} \bigg[ G_2 (\omega) \int_t^\infty G_2 (\tau) \, e^{i \omega \tau} \bigg] \label{q2} \\
&&   \approx - \frac{2
\sin(\alpha\pi)\Gamma^2(1+\alpha)\sin^2(\frac{\pi
\alpha}{2})}{M\Omega\pi(1+2\alpha)}\Delta^2(\Omega t)^{-2\alpha-1}\, .
\label{q2_approx}
 \eea
In Figs.~(\ref{qq_2}) and (\ref{qq_3}), we numerically check the
asymptotic behavior of $\langle \hat q^2 (t) \rangle_{F,{\rm
asy}}$ for the narrow and broad resonance cases respectively. This
is done by using the small $\omega$ approximate of $G_H$
in~(\ref{G_0_H_approx}) and  $G_2(\omega)$ in (\ref{G2omega}).
The numerics shows that the initial oscillation damps out and afterwards the
position uncertainty settles to the power law relaxation at a rate
$t^{-2\alpha -1}$ as we find in (\ref{q2_approx}). As
compared with $\langle q^2 (t) \rangle_I$
in~(\ref{q2M_late_time}) , $\langle q^2 (t) \rangle_{F,{\rm asy}}$
dominates  at late times, and to sum up, we have
 \be \label{q2_late_time} \langle q^2 (t)
\rangle =\langle q^2 (\infty) \rangle_{F}- \frac{2
\sin(\alpha\pi)\Gamma^2(1+\alpha)\sin^2(\frac{\pi
\alpha}{2})}{M\Omega\pi(1+2\alpha)}\Delta^2(\Omega
t)^{-2\alpha-1}+{\cal O}((\Omega t)^{-2\alpha-2}) \, .
  \ee

Using the operator relation $\hat p(t)=M \frac{d}{dt} \hat q(t)$,
the cross correlation between the position and the momentum can be
found by simply taking the time derivative of $\langle \hat q^2
(t) \rangle$, and at late times, is given by
  \be \langle q(t)
p(t) + p(t) q(t)\rangle\simeq\frac{2}{\pi}
\sin(\alpha\pi)\Gamma^2(1+\alpha)\sin^2(\frac{\pi
\alpha}{2})\Delta^2(\Omega t)^{-2\alpha-2}+{\cal O}((\Omega
t)^{-2\alpha-3})
  \label{crosscor_late_time}
  \ee
with the vanishing saturated value. Following the same derivation,
the late-time behavior of the momentum uncertainty is obtained as
   \bea
   \langle \hat p^2 (t) \rangle &&
   =m^2 \dot{G}^2_1 (t) G_1 (t) \langle \hat q^2 (0) \rangle + \dot{G^2_2} (t)  \langle \hat p^2 (0) \rangle + M \dot{G}_1 (t) \dot{G}_2 (t) \langle \hat q(0)\hat p(0)+\hat p(0) \hat q (0) \rangle  \nonumber\\
   && \quad\quad  \quad +
 \frac{1}{ 2} \int_0^t  d \tau \int_0^t  d \tau' \dot{G}_2 ( t-\tau) \dot{G}_2 (t-\tau') \langle \big\{ \hat \eta (\tau), \hat \eta (\tau') \big\} \rangle       \nonumber\\
&& \simeq\langle p^2 (\infty) \rangle_{F} -\frac{2\Omega
M\sin(\alpha\pi)}{\pi(3+2\alpha)}\Gamma(1+\alpha)\Gamma(3+\alpha)\sin^2(\frac{\pi\alpha}{2})\Delta^2(\Omega
t)^{-2\alpha-3} +{\cal O}((\Omega t)^{-2\alpha-4})\, .\nonumber\\
  &&
\label{p2_late_time}
    \eea
where $\langle \hat p^2 (\infty) \rangle_{F} $ turns out to be finite
as $\delta \rightarrow 0$ for all $\alpha>1$ and is given by
  \be
 \label{pp_s}
\langle \hat p^2 (\infty)
\rangle_{F,\delta=0}=\frac{M^2z^2r_b^{z(3-\alpha)}}{\pi
  T_{n+1}S_n}\int_0^1 dx \frac{x^2g(x)}{|Bx^2+x f(x)|^2} \, .
 \ee
Substituting~(\ref{q2_late_time}),(\ref{crosscor_late_time}), and
(\ref{p2_late_time}) into the definition of the $w(t)$
function~(\ref{w}), we then obtain the Von-Neumann entanglement
entropy~(\ref{S_vn}).  We can similarly express the entanglement
entropy as the saturated term and the term vanishes as
$t\rightarrow\infty$
  \be
   S (t) = S (\infty)+ S_{\rm asy} (t)
\, .
  \ee

To the leading order in the small-$\delta$ limit, the saturated
value of the entanglement entropy  can be found from  $\langle
\hat q^2 (\infty) \rangle_{F}$ in~(\ref{q2_1}), (\ref{q2_s2}) and
$\langle \hat p^2 (\infty) \rangle_{F}$ in~(\ref{pp_s}). We find
that the saturated entanglement entropy has three qualitatively
different behaviors for the following different ranges of
$\alpha$,
 \begin{align}
 \label{s_s}
 S (\infty) & \simeq \begin{cases}
                                      \displaystyle \,\bigg(1-\frac{1}{\alpha}\bigg)\ln \bigg(\frac{r_b^z}{\Omega}\bigg) \, , \quad  1< \alpha <2 \, ;& \vspace{9pt}\\
                                    \displaystyle \,\bigg(\frac{3}{2}-\frac{\alpha}{2}\bigg)\ln \bigg(\frac{r_b^z}{\Omega}\bigg)\, , \quad  2< \alpha <3 \, ;& \vspace{9pt}\\
                                    \displaystyle  \,\bigg(\sqrt{W (\alpha)}+\frac{1}{2} \bigg) \ln \bigg( \sqrt{W (\alpha)} +\frac{1}{2} \bigg) - \bigg(\sqrt{W(\alpha)}-\frac{1}{2} \bigg) \ln \bigg( \sqrt{W (\alpha)} -\frac{1}{2}\bigg)
                                    \, ,   \quad \alpha >3
                                      \,&
                                \end{cases}
\end{align}
where
  \be
  W(\alpha)=\frac{z\rho_n^2r_b^{2z(2-\alpha)}}{\pi^2T_{n+1}^2} \int_0^1 dx \frac{g(x)}{|Bx^2+x f(x)|^2}\int_0^1 dx \frac{x^2g(x)}{|Bx^2+x f(x)|^2} \, .
  \ee
Here $W(\alpha)$ is the function $w(t\rightarrow\infty)$ in the
case $\alpha>3$. For $\alpha>3$, the saturated value is finite and
independent of the UV and IR cutoffs. In the unit that
$\rho_n\simeq L^{-n-1}$, we find $W(\alpha)\simeq
\big(\frac{l_p}{L} \big)^{2z(\alpha-1)} \big(\frac{L}{r_b}
\big)^{2z(\alpha-2 )}$, which is naturally very small. $W(\alpha)$
could be of the order one and consistent with the minimum
uncertainty relation as $L \sim r_b \sim l_p$, but this leads to
the breakdown of the classical gravity limit. Quantum gravity
effect seems to become important for $\alpha >3$ and this deserves
further study. Here we mainly focus on $1<\alpha <3$ but
$\alpha\neq 2$ with the same range of $\alpha$ being considered in
\cite{Tong_12} in the case of  the particle ($n=0$). The saturated
value of the Von Newmann entropy can be considered as a
measurement of number of the degrees of freedom available to the
system from the environment. Thus, for $1<\alpha<2$, our result
indicates that the number of the effective degrees of freedom
increases with increasing $\alpha$, while for $2<\alpha<3$, its
number decreases with increasing $\alpha$. The maximum saturated
entropy occurs when $\alpha$ approaches $2$, namely $z=n+2$, which
happens as the mass term changes from an irrelevant operator for
$\alpha <2$ to a relevant operator for $\alpha >2$. This is one of
the main results in this paper.

The asymptotic
behavior of the entanglement entropy toward the saturation is
determined by $\langle \hat q^2 (t) \rangle_{F, \rm asy} $, which
decays most slowly comparing to other terms in $w(t)$ at the late
time with the behavior as
 \be
 S_{\rm asy} (t) \simeq \frac{\langle \hat q^2 (t) \rangle_{F, \rm asy}}{\langle \hat q^2 (\infty)
\rangle_{F}}= - \frac{2
\sin(\alpha\pi)\Gamma^2(1+\alpha)\sin^2(\frac{\pi
\alpha}{2})}{m\Omega\pi(1+2\alpha)\langle \hat q^2 (\infty)
\rangle_{F}}\Delta^2(\Omega t)^{-2\alpha-1}\, .
 \ee
The entanglement entropy shows the power-law relaxation at the
late time.  Moreover, the larger value of $z$ and smaller value of $n$
lead to smaller value of $\alpha$ and the smaller saturation rate.
This is consistent with the field theory picture, where the
relaxation is due to the energy flow from UV to IR degrees of
freedom. The quantum critical theory in $d-1$ spacial
dimension with the dispersion relation $E\propto k^z$ for the
effective excitations has density of the states $\rho(E) \propto
E^{-1+(d-1)/z}$~\cite{Tong_12}. Since we consider the probed
$n$-dimensional mirror, the effective density of the states probed
by the mirror would be $\rho_e(E) \propto E^{-1+n/z}$. Thus the
number of the modes decreases with the increase in $z$ or
the decrease in $n$, which leads to the slower relaxation rate.
This is a remarkable result. The proposed holographic model gives
a very natural explanation to the relaxation rate that certainly
deserves an experimental test.

 \section{Comparison with the environment of  relativistic free field theory} \label{sec5}
\begin{figure}
\centering \scalebox{0.5}{\includegraphics[trim=2cm 4cm 2cm
2cm,clip]{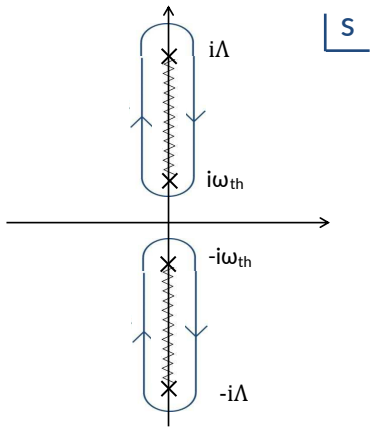}} \caption{The contour for the inverse
Laplace transform used to compute $G_{\phi,1}(t)$ and
$G_{\phi,2}(t)$ in a bath of the relativistic free field. There exist the cuts but the
poles are on other Riemann sheets.} \label{contour_phi}
\end{figure}

The environment-induced effects on the system, for example, a
point charge coupled to quantized electromagnetic
fields~\cite{hsiang} and a 2-dimensional moving mirror in the
environmental quantum free fields~\cite{wu} are to be summarized
in this section in the paradigm of quantum Brownian
motion~\cite{boyan,brbook,hubook}. In the linear response
approximation, the equation  of motion of either point charge or
moving mirror can be cast into the Langevin equation in the
form~(\ref{langevin}). Then, the self-energy in general can be
effectively expressed in terms of the spectral density,
$J(\omega)$ of the bath field denoted by the $\phi$ field as
 \be
 \Sigma_{\phi} (s) =-\frac{2}{\pi} \int_0^\infty d\omega J(\omega) \frac{ \omega}{ s^2+\omega^2} \, , \label{Sigma_phi}
 \ee
where $ J(\omega) \propto \omega^{\beta}$. For example,  a charged
point particle coupled to the quantized electromagnetic field
gives $\beta=3$ and a 2-dimensional mirror moving in the medium of
a relativistic free scalar field corresponds to $\beta=5$. The
solution of the Langevin equation can similarly be constructed
from the fundamental solutions $G_{\phi, 1} (t)$ and $G_{\phi, 2}
(t)$, obeying the homogeneous part of~(\ref{langevin}) with the
initial conditions~(\ref{G1ini}) and (\ref{G2ini}). The Laplace
transforms of the solutions are the same as the ones given
by~(\ref{Gs}), with the self energy replaced by
$\Sigma_{\phi}(s)$. The inverse Laplace transforms depend on the
analytical structure of $\Sigma_{\phi}(s)$. It can be seen
from~(\ref{Sigma_phi}) that the self-energy $\Sigma_{\phi} (s) $
can have a branch-cut along the imaginary $s$ axis. The real part
and the imaginary part of $\Sigma_{\phi}(s)$ can be constructed by
letting $s=i \omega \pm \epsilon$ in (\ref{Sigma_phi}) where the
resulting integral in the small $\epsilon$ limit is given by the
principal value. Then we have,
  \be
\Sigma_\phi (s=i \omega \pm  \epsilon)= { {\rm Re}} \Sigma_\phi
(\omega) \pm i { {\rm Im}} \Sigma_\phi (\omega) \, , \ee for
$\omega >0$ where \bea
{{\rm  Re}} \Sigma_{\phi} (\omega) &&= \frac{2}{\pi } \int_{\omega_{\rm th}}^{\Lambda} d\omega' \frac{J (\omega') \, \omega'}{\omega^2-\omega'^2} \, , \label{SigmaphiRe}\\
{ {\rm Im}} \Sigma_{\phi} (\omega)   &&= \frac{2}{\pi } {\rm sgn}
(\omega) \int_{\omega_{\rm th}}^{\Lambda} \, d\omega' J(\omega')
\, \omega' \delta(\omega^2-\omega'^2)  \, . \label{SigmaphiIm}
\eea
Here we impose a threshold energy $\omega_{\rm th}$ and a
UV-cutoff $\Lambda$ in the definition of the self-energy. The
branch cut then splits into two segments $(i\omega_{\rm th},
i\Lambda$), and also $(-i\Lambda, -i\omega_{\rm th})$ (See
Fig.(\ref{contour_phi})).

As the Brownian particle is trapped in a harmonic potential with
the frequency $\Omega$, the location of the poles in the inverse
Laplace transform can be found by solving the pole equation
similar to the one in~(\ref{poleeq}). Following \cite{boyan}, it
is convenient to define the renormalized self-energy and
frequency,
 \be
 \Sigma_{\phi, R} (s) =\Sigma_\phi (s) -\Sigma_\phi (0) \, ; \quad \quad\quad \Omega_R^2 =\Omega^2 +\Sigma_\phi (0) \, .
 \ee
The subscript $R$ will be omitted in the following discussion for
simplifying the notations. In the case that the interaction between the system
and environment are weak and the environmental fields are free
such as in \cite{hsiang,wu}, the positions of the poles can be
determined perturbatively with the corrections from environmental
fields.  As $\Omega<\omega_{\rm th}$, the
locations of the poles, $s_p$ is on the imaginary axis, leading to
the late time oscillatory behavior. As $\omega_{\rm
th}<\Omega<\Lambda$, the free particle pole is on the branch-cut,
and as can be seen from (\ref{SigmaphiIm}), the perturbative
effects from the environment shift the free particle poles to the
second Riemann sheet, then becoming the resonances \cite{boyan}.
This gives a clear mechanism that the decay of the system  (a
resonance) is due to the existence of the open channel that allows
the energy transfer to the environment from the system.
 Thus, as $\omega_{\rm th} <\Omega<\Lambda$, the inverse Laplace transform can be carried out by evaluating on the branch cuts, which gives
 \be \label{G_phi_2}
 G_{\phi,2} (t) = \int_{0}^{\Lambda} d \omega \,  \frac{ 4 \, {\rm Im} \Sigma_{\phi} (\omega)}{[ \omega^2+\Omega^2 + {\rm Re} \Sigma_{\phi} (\omega)]^2 + [{\rm Im} \Sigma_{\phi}(\omega)]^2}\, \sin (\omega t) \, .
 \ee
This integral over the $\omega$ in the limit of the weak coupling
between the system and bath exhibits a Breit-Wigner feature of the
narrow resonance. The width of the resonance is determined by the
imaginary part of the self-energy and its peak is located at the
shifted resonance frequency around $\Omega$. For the intermediate
times ( $1/\Omega \ll t \sim1/\Gamma_{\phi}$), the real-time of
$G_{\phi,2} (t)$ can be approximated by
 \be
  G_{\phi,2} (t) \simeq Z_\phi  \cos[ \Omega_\phi t + \theta_\phi] \,  e^{-\Gamma_\phi t} \, ,
  \ee
  where
  \be
  \Omega_\phi \simeq \Omega+ \frac{  {\rm Re} \Sigma_\phi (\Omega_\phi)}{ 2 \Omega_\phi} \, \quad\quad
  \Gamma_\phi \simeq \frac{ Z_\phi {\rm Im} \Sigma_\phi (\Omega_\phi)}{ 2 \Omega_\phi} \, .
  \ee
  with $ \Omega \gg \Gamma_{\phi}$ obtained perturbatively.
 Moreover, the phase shift $\theta_\phi$ and $Z_\phi$ are
 \be
   Z_\phi \simeq \bigg[ 1-\frac{\partial {\rm Re} \Sigma_\phi (\Omega_\phi)}{\partial \Omega_\phi^2} \bigg]^{-1} \, ,\quad\quad
   \theta_\phi \simeq Z_\phi \frac{\partial {\rm Im} \Sigma_\phi (\Omega_\phi)}{\partial \Omega_\phi^2} \, ,
  \ee
which are all perturbatively small. Finally, in the much later
time when $t \gg 1/\Gamma_{\phi}$, $G_{\phi,2} (t)$ decays in a
power law in time as $1/t^{\beta+1}$ determined by the small
$\omega$ behavior of the integrand in (\ref{G_phi_2}) where the
imaginary part of the self-energy is in (\ref{SigmaphiIm}) with
the spectral density $ J(\omega) \propto \omega^{\beta}$. The
fundamental function $G_{\phi, 1}(t)$ shows the similar relaxation
behavior. So, in the end, the entanglement entropy constructed
from these fundamental functions $G_{\phi 1}$ and $G_{\phi,2}$
will also exhibits the power law relaxation toward the saturation.

As a comparison, the self-energy of the system with the effects
from the strongly coupled quantum critical fields shows the
damping term of the form, $\omega^{\alpha}$ for non-integer
$\alpha$ whereas for a relativistic field theories the powers of
$\omega$ in the damping term is an integer as expected. Although
the positions of the cuts in a complex $s$ plane are very
different in both cases, at late times the system relaxes in a
power law with the inverse of the powers of time $t$ due to the
cut contributions. In particular, the existence of the cuts on the
imaginary part of the self-energy for free relativistic fields can
be explained by the energy transfer from the system to the bath
via creating quantum excitations of the environment. However for
quantum critical fields the cuts arise from the damping term with
peculiar non-integer powers of $\omega$ dependence, giving the
relaxation rate that also can be realized as  the energy flow to
the environment by counting the number of available modes of
environmental quantum critical fields.

Moreover, the holographic approach allows us to include the
non-perturbative effects from the environment. Thus, we find the
complex-valued pole solutions for either broad or narrow
resonances. However, the effects from relativistic free fields for
a weak coupling between the system and the environment are treated
perturbatively, giving a reliable result only on the narrow
resonance with relatively small corrections from the environment.
They all lead to the exponential decay in the intermediate time
scales. To summarize, in both cases, the effects of the
environment fields on the system share the same feature that the
system decays exponentially from the initial state during the
intermediate times given by the width of the resonance, and turns
to the power law relaxation determined by the small $\omega$
behavior of the self-energy.

\section{Summary and Outlook}\label{sec6}
The main goal of this work is to understand the time evolution of
the entanglement entropy between the $d$-dimensional strongly
coupled quantum critical field with a dynamical exponent $z$ at
zero temperature and a $n$-dimensional mirror using the method of
holography. The dual description is a $n+1$-dimensional probe
brane moving in $d+1$-dimensional Lifshitz geometry. The dynamics
of the mirror can be realized from the motion of the brane at the
boundary of the bulk. The interaction between the system and the
environment may result in the loss of the information of the
system, which can be measured by the von Neumann entropy,
$S=-Tr\rho_r\ln\rho_r$, computed from the reduced density matrix
$\rho_r$ of the system. In the linear response approximation, we
construct the holographic influence functional by tracing out the
environment's degrees of freedom. Then, the stochastic effective
action with the noise term manifested from quantum fluctuations of
the environment field is obtained, from which the associated
Heisenberg equations of the mirror with effects from the
environment are  derived. We consider the environment  at zero
temperature, and prepare an initial density matrix of the mirror
trapped by a harmonic potential in its ground state. Two sources
of the averages need to be dealt with: one is the average over the
intrinsic quantum uncertainty of the mirror  and the other is the
stochastic average induced by the quantum fluctuations of the
environment. When turning on the interaction of the system and the
environment, the entanglement entropy between them in the linear
response approximation can be found straightforwardly from the
position and the momentum uncertainties as well as the expectation
values of position-momentum cross correlations of the system by
solving the Heisenberg equations. The self-energy of the mirror
due to the effect of the environment not only gives the
corrections to the poles but also shows the existence of the cuts
in terms of the Laplace transformed variable $s$. We find that for
$1<\alpha<3$ but $\alpha\neq 2$, the entanglement entropy at the
late times follows a power law relaxation at a rate
$1/t^{2\alpha+1}$ to the saturation, due to the cut contributions.
We also find that the saturated values of the entanglement entropy
show two qualitatively different behaviors in the regions
$1<\alpha<2$ and $2<\alpha<3$. Moreover its relaxation dynamics
can be explained by counting the number of the modes of the
environments from the field theory perspective. We then compare
with the system in the bath of relativistic free fields. The
relaxation dynamics of the entanglement entropy in that case
follows the similar power-law relaxation at the late times, where
the existence of the cut has a clear explanation from the transfer
of energy from the system to the environment.

The immediate extension of our work is to study the dynamics of
relaxation and thermalization of the mirror coupled to quantum
critical fields at finite temperature. Another  extension of the
current work  is to consider two quantum systems coupled to one
strongly coupled quantum field. In particular, we may explore the
development of their quantum entanglement through the interaction
with the common environment field. On the one hand, the
environmental effects will induce quantum decoherence and
disentanglement. On the other hand, the environment as suitably
prepared or attuned to,  can also assist in maintaining or even
generating entanglement. To do so, one needs to extend the current
holographic setup to include two objects moving in the asymptotic
Lifshitz background.

Additionally, it will be of interest to compare our results with
the time dependent entanglement entropy obtained based upon
Ryu-Takayanagi conjecture~\cite{Ryu,Hubeny_07}. In
\cite{Hartman_13} and \cite{Liu_13}, the relaxation of the
entanglement entropy between two geometric regions in strongly
coupled fields with $z=1$ was studied by preparing a
non-stationary initial state and applying a global quench
respectively. Their results show generic power law relaxation in
the late times, although the detailed relaxation rate may depend
on the shape of the regions. In their studies, the entanglement
entropy saturates at some finite times, also determined by the
geometry of the regions. In our case, the entanglement entropy
shows the similar power law relaxation, but it saturates
asymptotically as the time goes to $t\rightarrow \infty$ instead.
This may due to the fact that the perturbation in their cases is a
global quench, whereas in our case the perturbation is local so it
takes infinite time to establish the entanglement to the whole
environment. The environment fields in their case are for quantum
critical field with $z=1$ ($\alpha$ is integer-valued) and also in
finite temperature.  It will be interesting to extend our study to
those cases and make the comparison. Moreover, in \cite{Fonda_14},
they studied the global quench in the Lifshitz background that
extended the work in \cite{Liu_13}. They found that in the case
that the entanglement boundary is a sphere, the late time
saturation rate of the entanglement entropy is independent of $z$,
but the early time power-law growth has the similar dependence on
$\alpha=1+\frac{2}{z}$ as in our case with local perturbation. The
detailed comparison deserves further study.

\begin{acknowledgments}
 This work was supported in part by the
Ministry of Science and Technology, Taiwan.
\end{acknowledgments}

\section{Brief summary of the holographic influence functional method}
 \label{zero order}
 Consider the Lifshitz black hole background with the metric
\be
  \label{lifshitz bh_a}
  ds^2=-r^{2z}f(r)dt^2+\frac{dr^2}{f(r)r^2}+r^2 dx_i dx_i \, ,
   \ee
where $f(r)\rightarrow1$ for $r\rightarrow\infty$ and $f(r)\simeq
c(r-r_h)$ near the black brane horizon $r_h$ with
$c=({d+z-1})/{r_h}$. With the same notations and assumptions as in
the main text, the DBI action for the $n+1$-dimensional probe
brane in the Lifshitz black hole for small perturbation $X^I$
around the stationary configuration is given by \be
  S^T_{DBI}
  \approx {\rm constant}- \frac{T_{n+1}}{2}\int dr \, dt \, dx_1 \, dx_2 \, ... \, dx_n \,
\bigg( r^{z+n+3} f(r) X'^{I} X'^{I}-
\frac{\dot{X}^{I}\dot{X}^{I}}{ { f(r) r^{z-n-1}}}\bigg) \, .
  \label{s_T_dbi_a}
  \ee
The equation of motion for $X^I$'s in the Fourier space,
$X^I_\omega (r)e^{-i\omega t}$ can then be derived as follows
 \be
  \label{mirror n with T_a}
 \frac{\partial}{\partial r}\biggl( r^{z+n+3}f(r)\frac{\partial }{\partial
 r}  X^I_\omega (r) \biggr)+\frac{\omega^2}{r^{z-n-1}f(r)} X^I_\omega(r) =0 \,
 .
  \ee
The solution can be expressed in terms of two linearly independent
solutions with the properties $\mathcal{X}_{\omega}(r)_{\substack{
   \propto \\
   r\rightarrow r_h
  }} e^{+i\omega r_*}$ and
$\mathcal{X}^{*}_{\omega}(r)_{\substack{
   \propto \\
   r\rightarrow r_h
  }} e^{-i\omega r_*}\, ,$
where $r^*=\int drf(r)^{-1}r^{-z-1}$, and the normalization
condition $\mathcal{X}_{\omega}(r_b)=1$. Since the different
components of $X^I_\omega$ are decoupled in the linearized
equation of motion, we may just focus on one of the directions
$X^I$ and denote it by $Q(t,r)$. As described in the main text we
introduce $Q^+(t,r_1)$ and $Q^-(t,r_2)$, which correspond to the
branes living in two outside regions in the maximally extended
Lifshitz black hole geometry. Following~\cite{Yeh_14},
 which is consistent with~\cite{Son_09, Son_02}, $Q^{\pm} (\omega,r)$ are then uniquely determined
with extra boundary conditions
 \be
 \label{bc_a} {q^{\pm}(t)=Q^{\pm}(t,r_{b})}\,,
 \ee
to be
    \bea \label{Q_pm_a}
    &&Q^+(\omega,r_1)=\frac{1}{1-e^{-\frac{\omega}{T}}} \bigg[ (q^-(\omega)- e^{-\frac{\omega}{T}}q^+(\omega))
    \mathcal{X}_{\omega}(r_1)+ ( q^+(\omega)-q^-(\omega))
 \mathcal{X}_{\omega}^*(r_1) \bigg]\, ,\nonumber\\
    &&Q^-(\omega,r_2)= \frac{1}{1-e^{-\frac{\omega}{T}}} \bigg[ (q^-(\omega)- e^{-\frac{\omega}{T}}q^+(\omega))
    \mathcal{X}_{\omega}(r_1)+ e^{-\frac{\omega}{T}} ( q^+(\omega)-q^-(\omega))
 \mathcal{X}_{\omega}^*(r_1) \bigg] \, .  \eea
where $q^{\pm}(\omega)$ is the Fourier transform of $q^{\pm}(t)$,
which will be identified with the mirror's position in the
close-time-path formalism. This solution is then substituted into
the classical action, we then obtain the holographic influence
functional
\begin{align} \label{h_influence_fun_a}
     F(q^+,q^-)&=S^{\rm on-shell}_{DBI }(Q^+)-  S^{\rm on-shell}_{DBI }(Q^-)\notag\\
     &=-T_{n+1}S_n r_b^{z+n+3}\int \frac{d\omega}{2\pi}\left(Q^+(-\omega,r_b)\partial_r Q^+(\omega,r_b)-Q^-(-\omega,r_b)\partial_r Q^-(\omega,r_b)\right)\nonumber\\
     &=-\int\frac{d\omega}{2\pi}\biggl\{q^+(-\omega)\biggl[i\operatorname{Re}G_R(\omega)-(1+2n_\omega)\operatorname{Im}G_R(\omega)\biggr]q^-(\omega)\biggr.\nonumber\\
     &\qquad\qquad\qquad+q^-(-\omega)\biggl[-i\operatorname{Re}G_R(\omega)-(1+2n_\omega)\operatorname{Im}G_R(\omega)\biggr]q^-(\omega)\nonumber\\
     &\qquad\qquad\qquad-q^+(-\omega)\biggl[-2n_\omega\,e^{\frac{\omega}{2T}}\operatorname{Im}G_R(\omega)\biggr]q^-(\omega)\nonumber\\
     &\qquad\qquad\qquad-\biggl.q^-(-\omega)\biggl[-2(1+n_\omega)\,e^{-\frac{\omega}{2T}}\operatorname{Im}G_R(\omega)\biggr]q^+(\omega)\biggr\}\,,
\end{align}
where $S_n$ is mirror's volume and $G_R^{(T)}(\omega)=T_{n+1} S_n
r_b^{z+n+3}\mathcal{X}_{-\omega}(r_b)\partial_r\mathcal{X}_{\omega}(r_b)$
is the retarded Green function at the finite temperature $T$, as
can be seen from (\ref{influencefun2}) and (\ref{G_HR}). In the
zero temperature limit, there is an exact expression for the
solution in (\ref{mirror n with T_a}), satisfying the desired
boudnary conditions,
  \be
  \label{mode_a}
  \mathcal{X}_{\omega}(r)=\frac{r_b^{\frac{z+n+2}2}}{r^{\frac{z+n+2}2}}\frac{H^{(1)}_{\frac{n+2}{2z}+\frac{1}{2}}(\frac{\omega}{zr^z})}{H^{(1)}_{\frac{n+2}{2z}+\frac{1}{2}}(\frac{\omega}{zr_b^z})}
  \, .
  \ee
Hence the zero-temperature retarded Green's function {for
$\omega>0$} can be found
 to be,
  \be
\label{G_R_a} G_R(\omega)=- T_{n+1} S_n \,  {\omega \,
r_b^{n+2}}\frac{H^{(1)}_{\frac{n+2}{2z}-\frac12}(\frac{\omega}{zr_b^z})}{H^{(1)}_{\frac{n+2}{2z}+\frac12}(\frac{\omega}{zr_b^z})}
\,.
  \ee
 The zero-temperature
Hadamard function for $\omega
>0$ can also be found using (\ref{influencefun2}) and
(\ref{G_HR}),
 \be \label{G_0_H_a}
 G_H(\omega)=\frac{2 z}{\pi} r_b^{n+2+z} \frac{T_{n+1} S_n} {J^2_{\frac{n+2}{2z}+\frac12}(\frac{\omega}{zr_b^z})+Y^2_{\frac{n+2}{2z}+\frac12}(\frac{\omega}{zr_b^z})}
  \, . \ee

\end{document}